\documentstyle[11pt,here]{article}
\input epsf
\def\mskip{\hspace{.05in}}
\def\mskipp{\hspace{.1in}}
\def\hs{\hspace{.25in}}
\def\href{\hspace{.2cm}}
\def\ns{\hspace{-.04in}}

\def\beq{\begin{equation}}
\def\eeq{\end{equation}}
\def\beqa{\begin{eqnarray}}
\def\eeqa{\end{eqnarray}}
\begin{document}
\baselineskip=24pt

\begin{center}
{\Large\bf NEGATIVE ENERGY MODES AND GRAVITATIONAL}

{\Large\bf INSTABILITY OF INTERPENETRATING FLUIDS}

\vspace{.2cm}
A. R. R. CASTI$,^{a}$ P. J. MORRISON$,^{b}$ AND E. A. SPIEGEL$^{c}$

${ }^{a}${\it Department of Applied Physics/}
\vspace{-.4cm}

{\it Division of Applied Mathematics}

\vspace{-.4cm}
{\it Columbia University}

\vspace{-.4cm}
{\it New York, New York 10027}

${ }^{b}${\it Department of Physics and Institute for Fusion Studies}

\vspace{-.4cm}
{\it The University of Texas at Austin}

\vspace{-.4cm}
{\it Austin, Texas 78712}

${ }^{c}${\it Department of Astronomy}

\vspace{-.4cm}
{\it Columbia University}

\vspace{-.4cm}
{\it New York, New York 10027}
\end{center}



\hspace{2.5cm}{\it A negative energy electron will have less energy the}

\vspace{-.4cm}
\hspace{2.5cm}{\it faster it moves and will have to absorb energy in} 

\vspace{-.4cm}
\hspace{2.5cm}{\it order to be brought to rest.  No particles of this nature} 

\vspace{-.4cm}
\hspace{2.5cm}{\it have ever been observed.} ------ P. A. M. Dirac, 1930

\vspace{.5cm}
\begin{abstract}
\baselineskip=20pt
\noindent
We study the longitudinal instabilities of two interpenetrating
fluids interacting only through gravity.  When one of the
constituents is of relatively low density,
it is possible to have a band of unstable wave numbers
well separated from those involved in the usual Jeans instability.
If the initial streaming is large enough, and there is no linear
instability, the indefinite sign of
the free energy has the possible consequence of
explosive interactions between positive and negative energy modes
in the nonlinear regime.  The effect of dissipation on the
negative energy modes is also examined.
\end{abstract}
\vspace{.5cm}
\eject
\section{BACKGROUND}
\label{sec-gravback}
\hs
It used to be thought that the stellar component of two spiral
galaxies would pass right through each other in the event of
a collision and that only the gaseous components would merge.
However, simulations over the past twenty years or so$^{1,2}$
have shown that the macroscopic energy of such a collision is quickly
converted into internal energy and that merger of the stellar systems
is a common natural outcome of a collision.  How is this conversion effected?

An answer to this may lie in the physics of streaming instabilities.
In the context of plasma physics, interpenetrating electron beams
produce the two-stream instability$^{3}$
whose gravitational analog has long been recognized, beginning
with the investigations of Sweet$^{4}$ and Lynden-Bell.$^{5}$
Of course, in the case of galaxy collisions,
which occur  quickly, the conventional two-stream instability may operate
too slowly to be effective.  However, it is known that even
when two streams of interacting particles are not linearly unstable, they
may collectively produce negative energy modes that lead to an explosive
nonlinear growth of perturbations for arbitrarily small disturbances.
There are well-developed criteria for the occurrence of this explosive growth
in plasma physics$^{6}$ and, 
as Lovelace {\it et al.}$^{7}$ have suggested
in their analysis of counter-rotating galaxies, we may expect something
analogous in the gravitational setting.  Our aim here is to briefly
develop this topic of negative energy modes for the case of gravitational
interaction in the expectation that the phenomena involved will be found
significant in a variety of astrophysical processes.

Even in the event of linear instability, the case of counterstreaming
populations is significantly different from the standard gravitational
instability, which occurs only for perturbation scales greater than the
Jeans length.$^{8}$  When there are two interpenetrating fluids
such as stars and gas, modes of arbitrary wavelength can be rendered
unstable.  Numerous investigators have reported on these issues.
Most of them (such as Ikeuchi et al.,$^{9}$ Fridman \& Polyachenko,$^{10}$
and Araki$^{11}$) focused primarily on the symmetric situation
of identical fluids in counterstreaming motion.  In that case one
finds that the spectrum of any instabilities arising from the relative
motion is wholly contained within the Jeans instability band, and
this blurs the distinction between the two processes.  This need not
be true when this symmetry is broken, and indeed not all authors
restricted themselves entirely to the symmetric case.  The present
venture into the asymmetric problem is intended to focus on the
possibility of well-separated instability bands, which has not been
elucidated in the gravitational context, as far as we are aware.

\section{EQUATIONS OF MOTION}
\label{sec-graveqns}
\hs
To see the problem in its simplest version, it is useful
to have a uniform medium as the unperturbed state.  Rather
than formulate the problem inconsistently to achieve this
end, as Jeans did, we prefer the Einstein device of introducing
a cosmological repulsion term.  In the Newtonian setting we
readily see how to redefine the gravitational potential so that,
instead of introducing such a repulsion term, we fill space with
a fluid of {\it negative} gravitational mass of density
$\rho_{_\Lambda}$.  As in the one-fluid plasma model, we treat
this density as a constant since its purpose is to allow a
gravitationally neutral background state.  One may also contemplate
the analog of the two-fluid plasma model in which this background
antigravitational fluid has its own dynamics, but we do not do
that here.  The two dynamically active fluids we consider are gravitationally
ordinary and polytropic.  Thus, the Poisson equation is written here
as,
\beq
\nabla^2 V=4\pi G\left(\rho_1+\rho_2-\rho_{_\Lambda}\right),
\label{eq:3Dgrav}
\eeq
where $V$ is the gravitational potential, $\rho_1$ and $\rho_2$ are
the source densities of the conventional fluids, and
$-\rho_{_\Lambda}$ is the cosmological background density.

The equations of motion for the two fluids ($j=1,2$) are,
\beqa
\rho_j\left(\partial_t+{\bf u}_j\cdot\nabla\right){\bf u}_j&=&
-\nabla p_j-\rho_j\nabla V \label{eq:3Dmom}\\
\partial_t \rho_j+\nabla\cdot\left(\rho_j{\bf u}_j\right)&=&0,
\label{eq:3Dmass}
\eeqa
where we do {\it not} sum over repeated indices.
Each fluid has a sound speed,
$c_{j}^{2}=${\Large{${\frac{\partial p_j}{\partial\rho_j}}$}},
and a Jeans wavenumber,
$k_{{Jj}}^{2}=${\Large${\frac{8\pi G{\widehat\rho_j}}{c_{j}^{2}}}$},
where the hat signifies the equilibrium value and an uncustomary
factor of $2$ appears in the definition of the Jeans wavenumbers.

We shall use natural units with $k_{_{{J1}}}^{-1}$ as the length
scale and $(k_{_{{J1}}}c_1)^{-1}$ as the time scale.  We further
simplify the description by considering only longitudinal motions
in one-dimension, so each single-component velocity field may be expressed
as the gradient of a velocity potential:  $u_j=\partial_x\phi_j$.
The fundamental equations (\ref{eq:3Dgrav})-(\ref{eq:3Dmass})
take the dimensionless form,
\beqa
M_{1}\partial_t\phi_1+{\frac{1}{2}}M_{1}^{2}(\partial_x\phi_1)^2+
{\frac{\rho_1^{\gamma_{_1}-1}}{\gamma_{_1}-1}}+V&=&B_{_1} \label{eq:bern1}\\
cM_{2}\partial_t\phi_2+{\frac{1}{2}}c^2 M_{2}^{2}(\partial_x\phi_2)^2+
{\frac{c^2\rho_2^{\gamma_{_2}-1}}{\gamma_{_2}-1}}+V&=&B_{_2} \label{eq:bern2}\\
\partial_t\rho_1+M_{1}\partial_x(\rho_1\partial_x\phi_1)&=&0
\label{eq:mass1}\\
\partial_t\rho_2+cM_{2}\partial_x(\rho_2\partial_x\phi_2)&=&0
\label{eq:mass2}\\
\partial_{x}^{2}V-{\frac{1}{2}}\rho_1-
{\frac{1}{2}}\beta\rho_2+{\frac{\rho_{_\Lambda}}{2\widehat\rho_1}}
&=&0\mskip,\label{eq:grav}
\eeqa
where the $M_{j}=${\Large${\frac{{\cal U}_j}{c_j}}$} are Mach numbers,
${\cal U}_j$ measures the initial streaming velocities,
$\beta=${\Large${\frac{\widehat\rho_2}{{\widehat\rho_1}}}$},
$c=${\Large${\frac{c_2}{c_1}}$}, and the Bernoulli constants $B_{_j}$
are chosen to balance the basic state.

\section{LINEAR THEORY}
\label{sec-gravlin}
\hs
We perturb from the state of uniform densities and
constant velocities by setting
$\phi_j=(-1)^{j+1}x+\delta\phi_j$, $\rho_j = 1+ \delta \rho_j$,
and $V={\widehat V}+\delta V$.  The density terms of the
Poisson equation (\ref{eq:grav}) combine to vanish and ${\widehat V}$ is
a constant.  Since the linearized equations are separable we may decompose the
perturbations into normal modes proportional to
$\exp(i\omega t-ikx)$ to find the dispersion relation,
\beq
\Gamma(\omega,k) = 1 +{\frac{1}{2\left[(\omega-kM_{1})^2-k^2\right]}}+
{\frac{\beta}{2\left[(\omega+ckM_{2})^2-c^2 k^2\right]}}\equiv
1+\Gamma_1+\Gamma_2=0. \label{eq:diagravic}
\eeq
The quantity $\Gamma$, which we call the diagravic function by analogy
with the dielectric function of electrodynamics, measures the
collective response of the fluid to disturbances in the gravitational
field and will serve to indicate the energy signature of any normal
mode (section \ref{sec-gravnonlin}).

For real $k$ the solutions of (\ref{eq:diagravic})
with complex $\omega$
correspond to instability; if $\omega$ is real, then
solutions of (\ref{eq:diagravic}) with complex $k$ can give rise to
wave amplification instability.  Here we analyze only the case of 
real $k$.  However, we have to deal with both the traditional Jeans instability
as well as the two-stream instability, the latter of which 
involves a sympathetic bunching of particles and is effective
for creating instability when the phase speed of the disturbance
conspires to create a resonance between different modes.

\subsection{\bf Symmetric Case}
\label{sec-symm}
\hs
If both fluids have the same basic properties ($c=1$ and $\beta=1$),
a frame exists in which $M_{_1}=M_{_2}=M$.
The dispersion relation then
simplifies into a manageable biquadratic with solutions,
\beq
\omega_{\pm}^{2}=
-{\frac{1}{2}}+k^2(M^2+1)\pm\sqrt{{\frac{1}{4}}-2k^2 M^2+4M^2 k^4}.
\label{eq:symroots}
\eeq

For $M=0$, we recover a simple version of the previously studied
two-fluid Jeans \linebreak problem.$^{10,12,13}$
We find $\omega_{+}^{2}=k^2$, corresponding to sound waves at all $k$,
and $\omega_{-}^{2}=k^2-1$, which is the conventional Jeans dispersion
relation.  The new
acoustic modes arise because the aggregate fluid now allows motions
unaffected by the gravitational field; for these modes the perturbed
gravitational potential is zero.

With relative velocity in the subsonic regime ($0<M<1$),
there is only a single unstable
mode that branches continuously from the Jeans mode at $M=0$.
This mode is unstable for all wavenumbers below a critical value
that approaches infinity as $M$ tends to unity from below (see Table 1).
To study this limit, we let $M^2=1-\alpha/k^2$
with $0<\alpha<1$.  As $k\rightarrow\infty$ we find the approximate
solution $\omega_{-}^{2}\sim -${\Large${\frac{\alpha(1-\alpha)}{4k^2}}$},
which reveals a weak instability at large $k$.  Thus,
weak relative streaming allows
gravitational instability at arbitrarily small wavelengths.  These
large-$k$ instabilities do not arise for Maxwellian velocity distributions 
within the context of the Vlasov equation.$^{9}$

For supersonic motion ($M>1$) the large-$k$ gravitational instability
is no longer present, but a new instability emerges that we call a
two-stream instability since it owes its presence to the energy
contained in the initial streaming motion.  
As $M$ ranges from $1$ to $\infty$, the critical wavenumber for instability
increases from $k_{{crit}}=1/2$ to $k_{{crit}}=\sqrt{2}/2$.

The upper half of figure \ref{f:quarticM2} shows that near
$k=0$ the two-stream modes are wholly contained within the Jeans band.
This fact coupled with the larger growth rates of the Jeans modes
has led some to believe that the two-stream instability is swamped
by the Jeans instability and is essentially unimportant.$^{11}$
As $k$ increases the two-stream and Jeans modes collapse upon
each other and together bifurcate into growing and damped oscillations.
At still larger wavenumbers all motions are stable, propagating
waves.  The critical wavenumbers below which growth is possible
at any Mach number are shown in Table 1.

\begin{center}
{\bf TABLE 1}\footnote{The value of $k_{{crit}}^{2}$ for the
  two-stream modes with $M\geq 1$ more accurately refers to
  the $k$ value at which the unstable two-stream and Jeans
  branches merge.}

\vspace{.1in}
\vspace{.3in}
\begin{tabular}{|c|c|c|c|c|} \hline
Mach Range & Mode Type & $k_{{crit}}^{2}$ & $\lim_{M\rightarrow 1}
 k_{{crit}}^{2}$ & $\lim_{M\rightarrow\infty} k_{{crit}}^{2}$\\ \hline
$0\leq M< 1$ & Jeans & {\Large ${\frac{1}{1-M^2}}$}& $\infty$ &
 Not Applicable \\ \hline
$1\leq M$ & Two-Stream & {\Large
  ${\frac{\sqrt{M^2-1}}{4M(M^2-1+M\sqrt{M^2-1})}}$}
 & ${\frac{1}{4}}$ & $0$\\ \hline
$1\leq M$ & Jeans & {\Large
  ${\frac{\sqrt{M^2-1}}{4M(1-M^2+M\sqrt{M^2-1})}}$}
 &${\frac{1}{4}}$ & ${\frac{1}{2}}$\\ \hline
\end{tabular}
\end{center}
\vspace{.5in}

\subsection{\bf Asymmetric Case}
\label{sec-asymm}
\begin{figure}[H]
\epsfxsize=4.5in
\centerline{\epsfbox{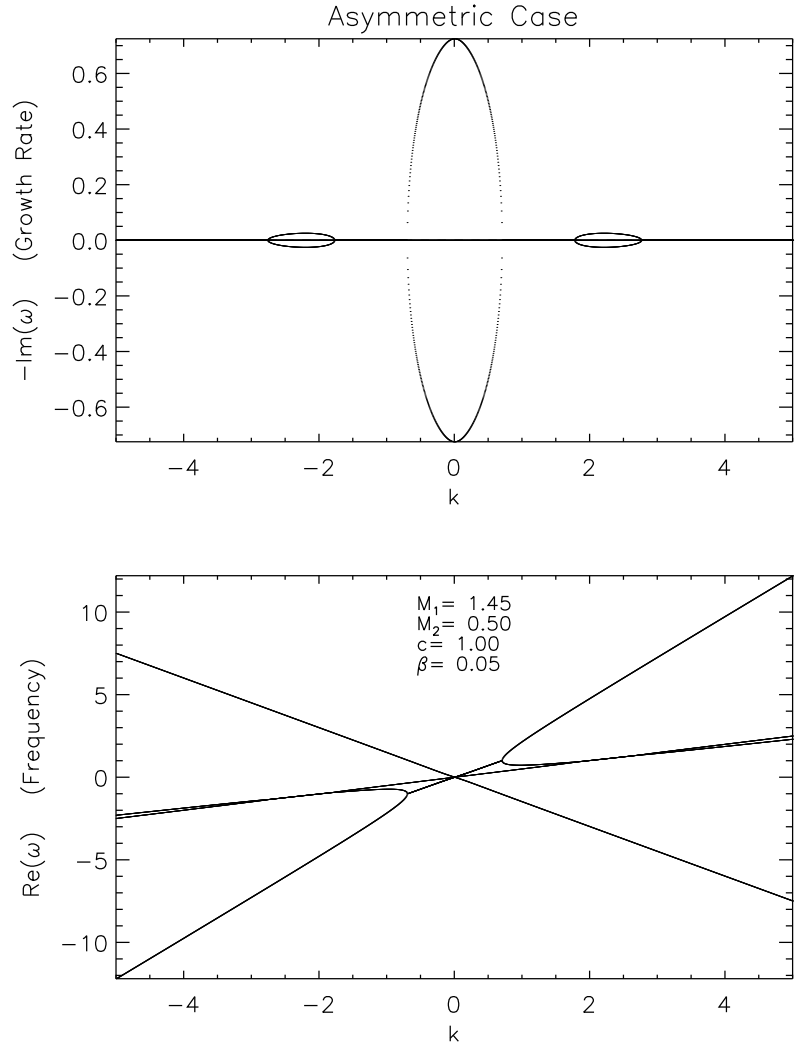}}
\caption{Asymmetric Case, $\beta=.05$, $c=1$, $M_1+M_2=1.95$.
Illustration of the pinched two-stream bubbles.
The frequencies corresponding
to the two-stream modes are degenerate along the entire $k-$band
of the bubble.  At the edge of the bubble the frequencies once again
separate.}
\label{f:wispyM1.45}
\end{figure}

\hs
When we relax the constraint of identical conditions in the two
fluids, one of the more interesting
consequences is the possibility of large wavenumber
two-stream instability bands well-separated from the Jeans
instabilities clustered at small $k$.  For illustration
we consider the effect of changing the initial relative streaming
$M_1+cM_2$ for fixed $\beta$ and $c$.  In fact, $\beta$ turns
out to be the crucial parameter in achieving the spectral separation;
variations in the sound speed ratio $c$ widens both bands together.
The distancing of a bubble of two-stream modes from the
Jeans band is illustrated in figure \ref{f:wispyM1.45}.

The large-$k$ two-stream modes can be explained qualitatively
by examining the separate pieces of the dispersion relation
(\ref{eq:diagravic}).  Since the streaming instability is
related to resonant motions, it
is revealing to examine the solutions to $1+\Gamma_1(\omega,k)=
1+\Gamma_2(\omega,k)=0$ in isolation and see where the
curves intersect.  The frequencies of these non-interacting modes are given by,
\beqa
\omega_1&=&kM_1\pm\sqrt{k^2-{\frac{1}{2}}} \label{eq:sepmode1}\\
\omega_2&=&-ckM_2\pm\sqrt{c^2 k^2-{\frac{\beta}{2}}} \label{eq:sepmode2}
\eeqa
When $\omega_1=\omega_2$, the assumed independent frequencies
match one another for wavenumbers satisfying,
\beq
 k(M_1+cM_2)=\left\{\begin{array}{l}
               \pm\left(\sqrt{k^2-{\frac{1}{2}}}+\sqrt{c^2
k^2-{\frac{\beta}{2}}}\right)\\
               \pm\left(\sqrt{k^2-{\frac{1}{2}}}-\sqrt{c^2
k^2-{\frac{\beta}{2}}}\right)
               \end{array} \right. \label{eq:rescon}
\eeq

In the symmetric case where $M_1=M_2\equiv M$, $c=\beta=1$, we see that two
of the
resonances are lost except in the irrelevant cases $M=0$ and $k=0$.
The other pair of possibilities, $kM=\pm\sqrt{k^2-{\frac{1}{2}}}$,
just restate the critical
wavenumber condition for what we know to be the modified Jeans
instability when $M<1$.  In the general case, we can expect another pair of
intersections that account for the two-stream bubbles of
figure \ref{f:wispyM1.45}.

\vspace{.5cm}
\section{NONLINEAR THEORY}
\label{sec-gravnonlin}
\vspace{.5cm}
\subsection{\bf Hamiltonian Formulation and Energy Signature}
\label{sec-ham}
\hs
The dynamical equations (\ref{eq:bern1})-(\ref{eq:grav}) derive from a
variational principle and a conserved Hamiltonian functional.
The variational formulation has the advantage of shedding
light on the relation between the energy content of the disturbances
and nonlinear stability.  Here we present results for
the symmetric case, though the formalism follows through for the
asymmetric case as well.

The Hamiltonian and associated equations are,
\beqa
H&=&{\frac{1}{M}}\sum_{j=1}^{2}\int_{0}^{L}
dx\left({\frac{M^2}{2}}\rho_j {\phi}_{jx}^{2}
  +\rho_j{U}_j -V_{x}^{2} \right),
\label{eq:ham}\\
\partial_t\phi_j&=&\left\{\phi_j,H\right\}=-{\frac{\delta
  H}{\delta\rho_j}}\label{eq:bernpoisson}\\
\partial_t\rho_j&=&\left\{\rho_j,H\right\}={\frac{\delta
  H}{\delta\phi_j}},\label{eq:masspoisson}
\eeqa
where ${U}_j (\rho_j)=${\Large ${\frac{\rho_{j}^{\gamma-1}}
{\gamma(\gamma-1)}}$} is the internal energy for the $j^{\rm th}$
fluid and the Poisson bracket is defined by,
\beq
\left\{F,G\right\}=\sum_{j=1}^{2}\int_{0}^{L}dx^{\prime}\left(
{\frac{\delta G(x)}{\delta\phi_{j}(x^\prime)}}{\frac{\delta
F(x)}{\delta\rho_{j}(x^\prime)}}
-{\frac{\delta F(x)}{\delta\phi_{j}(x^\prime)}}{\frac{\delta
G(x)}{\delta\rho_{j}(x^\prime)}}
\right).
\label{eq:poissonbracket}
\eeq
For definiteness we have chosen a box-geometry of length $L$.

The energy content of a particular mode when the perturbation
amplitude is small is given by
the second variation of $H$ evaluated at equilibrium (the free
energy).  After some calculation this is seen to be,
\beq
\delta^2 H={\frac{1}{2M}}\int_{0}^{L}dx\left(M^2\left[\delta\phi_{{1x}}^2+
\delta\phi_{{2x}}^2\right]+2M^2\left[\delta\rho_1 \delta\phi_{{1x}}
-\delta\rho_2 \delta\phi_{{2x}}\right]+\delta\rho_{1}^{2}
+\delta\rho_2^2-2\delta V_{x}^{2}\right).
\label{eq:freeenergy}
\eeq
It may be verified that this functional is conserved by the
equations of motion.  Suppose we now insert into (\ref{eq:freeenergy})
an eigenfunction corresponding to a single stable mode with
Im($\omega$)$=0$.  Employing overbars to denote eigenvector components
and $*$ for complex conjugation, we write,
\beq
\delta V={\overline{V}}e^{i(\omega t-kx)}+{\overline{V}_j^*}
e^{-i(\omega^{*} t-kx)}, \label{eq:rhoefunk}
\eeq
and similarly for the other perturbation variables.  Upon effecting
the integrations, we can make use of the dispersion relation,
\beq
\Gamma(\omega,k)=1+{\frac{1}{2\left[(\omega -kM)^2 -k^2\right]}}+
{\frac{1}{2\left[(\omega +kM)^2
-k^2\right]}}=0\mskip, \label{eq:nondimdia}
\eeq
to express the modal free energy in the compact form,
\beq
\delta^2 H=-2Lk^2\mid{\overline V}\mid^2
\omega{\frac{\partial\Gamma}{\partial\omega}}.
\label{eq:freediagravic}
\eeq

Wherever {\large$\omega{\frac{\partial\Gamma}{\partial\omega}}$}$<0$
a positive energy mode (PEM) is implied by (\ref{eq:freediagravic}),
while the condition {\large$\omega{
\frac{\partial\Gamma}{\partial\omega}}$}$>0$
defines a negative energy mode (NEM).  This possibility of modes of
either signature has been elucidated in the plasma physics literature$^{14}$
and a gravitational analog was suggested
by Lovelace et al.$^{7}$ in the context of thin, counter-rotating
stellar disks.

\begin{figure}[H]
\epsfxsize=4.5in
\centerline{\epsfbox{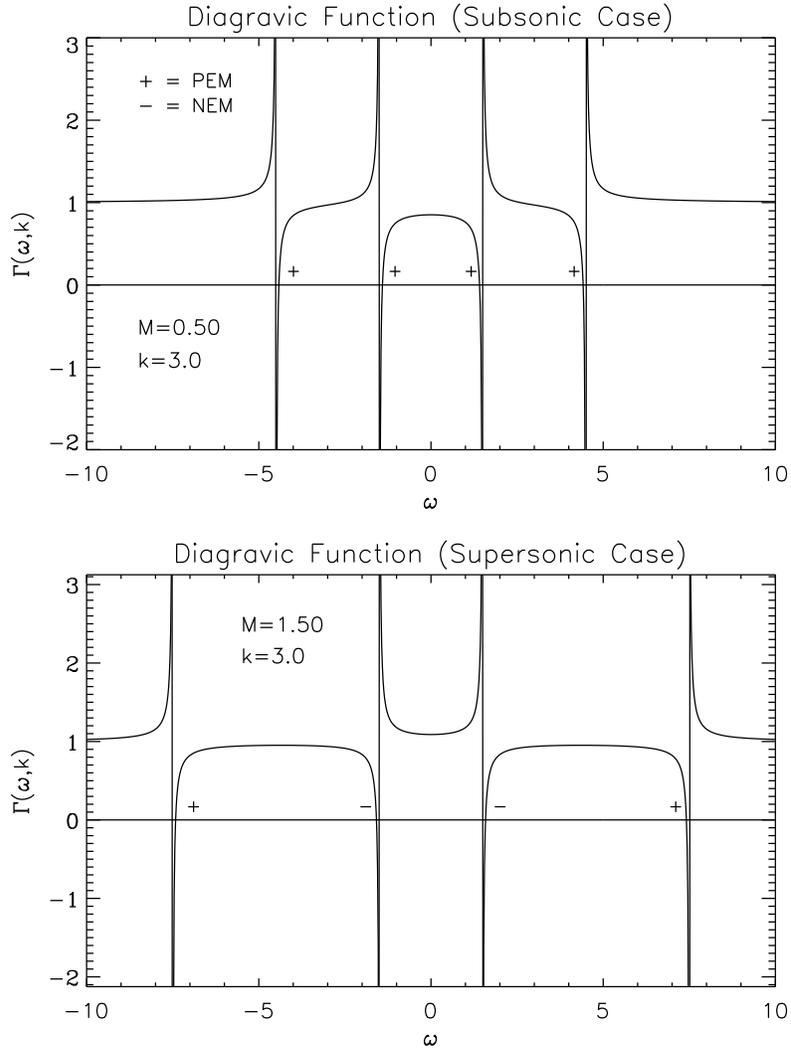}}
\caption{Diagravic function in the symmetric case for linearly stable
modes with $k=3$.  A stable mode exists wherever $\Gamma$
crosses the $\omega$-axis.  The positive
energy modes are indicated with a ``$+$'' and the negative energy
modes with a ``$-$''.}
\label{f:diagravic_SymmetricCase}
\end{figure}

Figure \ref{f:diagravic_SymmetricCase} shows the diagravic function for
both a subsonic and a supersonic case of stable modes with $k=3$.
In the subsonic regime, we find the Hamiltonian to be positive
definite near equilibrium since
{\large$\omega{\frac{\partial\Gamma}{\partial\omega}}$}$<0$ at every
crossing of $\Gamma$ on the $\omega$-axis.  Right at the border
of supersonic streaming ($M=1$), concomitantly with the appearance
of the two-stream instability, the $\Gamma$ curves undergo a
topological transition that allows the coexistence of positive
and negative energy modes.  The NEMs are {\it slow modes} in that
they have smaller frequencies than their PEM counterparts.  This
is the typical situation; $\omega$ must pass through zero if
the energy signature changes.$^{15}$  We expect from
the precedents of plasma physics$^{6}$ that the simultaneous presence
of positive and negative energy modes has dramatic consequences
on the nonlinear stability of the system.

\subsection{\bf Reduction to Action Angle Variables}
\label{sec-canon}
\hs
In the rest of this section we will concentrate on the nonlinear
interactions between linearly stable modes in the symmetric
problem (see figure \ref{f:diagravic_SymmetricCase}).  From a physical
standpoint, attention is focused on situations where the disturbances
are of sufficiently small scale so that the Jeans instability
can be ignored, though there are no
compelling reasons why this ought to be the case.  We will further
assume supersonic motion in order to examine the interaction of
positive and negative energy modes, a situation we expect to be the
most interesting.  Under these assumptions the
equations of motion achieve their simplest
form in action-angle coordinates that we now develop.

First we Fourier transform the field variables:
\beq
\rho_j=\sum_{m=-\infty}^{\infty}\rho_m^{(j)}(t)e^{ik_m x}
\mskipp,\mskipp \rho_{{-m}}^{(j)}={\rho_{{m}}^{(j)}}^*\mskipp,\mskipp
k_m={\frac{2\pi m}{L}}\mskip.\label{eq:fourier}
\eeq
We can then write the free energy in terms of real variables as,
\beq
\delta^2 H={\frac{1}{2}}\sum_{m=1}^{\infty}\left({\bf q}^{T}
{\bf A}{\bf q}+{\bf p}^{T}{\bf B}{\bf p}\right),
\label{eq:freeenergypq}
\eeq
where ${\bf q}\equiv(q_1,q_2,q_3,q_4)^{T}$,
${\bf p}\equiv(p_1,p_2,p_3,p_4)^{T}$, are linear combinations of the
complex modal amplitudes, ${\bf A}$ and ${\bf B}$
are symmetric matrices (given in Casti$^{16}$), and
the ``$T$'' indicates transpose.

Defining the configuration variables ${\bf z}=(q_1,\ldots,p_4)^{T}$,
we recast the linearized equations in the form,
\beq
{\frac{d{\bf z}}{dt}}={\bf J}\nabla_{{\bf z}}{\delta^2 H}\equiv
{\cal\bf L}{\bf z}\mskipp\mskipp,\mskipp\mskipp
{\cal\bf L}=\left(\begin{array}{clcr} 0 & {\bf B} \\ -{\bf A} & 0
\end{array}\right)
\mskipp\mskipp,\mskipp\mskipp
{\bf J}=\left(\begin{array}{clcr} 0 &{\bf I} \\ -{\bf I} & 0
\end{array}\right),
\label{eq:Hameqns}
\eeq
where ${\bf J}$ is the canonical $8\times 8$ cosymplectic form.  The
next order of business is to construct a symplectic transformation
that puts $\delta^2 H$ in its normal form.$^{17,18}$
This can be achieved by writing ${\bf z}={\bf S}{\bf Z}$, where
the matrix ${\bf S}$ consists of suitably ordered eigenvectors
of ${\bf L}$ satisfying the symplectic condition,
${\bf S}^{T}{\bf J}{\bf S}={\bf J}$.  After a final
transformation to action-angle coordinates, the free energy
expression (\ref{eq:freeenergypq}) becomes a superposition of harmonic
oscillators,
\beq
\delta^2 H=\sum_{m=1}^{\infty}\left(\omega_+ J_1+\omega_+ J_2
-\omega_- J_3-\omega_- J_4\right).
\label{eq:freeaa}
\eeq
The free energy is thus manifestly composed of two pairs each of positive
and negative energy modes.

\subsection{\bf Three-Wave Resonance and Explosive Growth}
\label{sec-3wave}
\hs
Energy conservation forbids nonlinear runaway growth if $H$ is
definite (Dirichlet's theorem), as is the case here for subsonic motion.
When the relative streaming is supersonic, however, interacting
PEMs and NEMs can circumvent this restriction since they contribute
energy of opposite sign.

We demonstrate the possibility of explosive growth with a three-wave
resonant interaction between two NEMs and one PEM.
Since the energy signature of a mode is not Galilean
invariant, the existence of a
reference frame in which all three modes have the same
signature implies nonlinear stability.  It may be shown
that there is no reference frame in which all three
modes have the same energy signature if the highest frequency
wave has opposite signature to that of the other two.$^{19}$
This provides a criterion for three-wave interactions leading to instability.

\begin{figure}[H]
\epsfxsize=4.5in
\centerline{\epsfbox{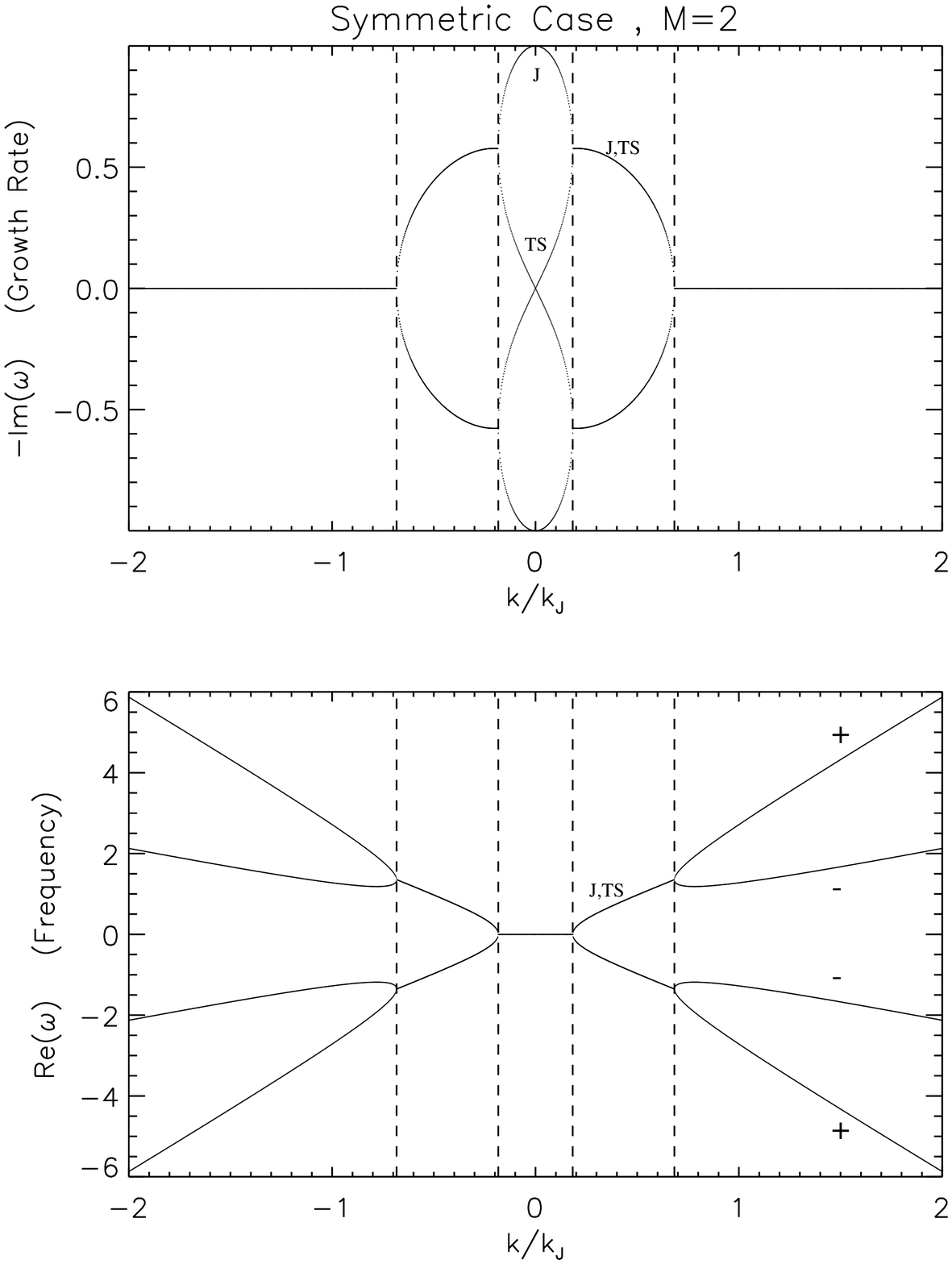}}
\caption{Dispersion curves for supersonic motion, $M=2$.  The
Jeans modes are indicated with a ``J'' and the
two-stream modes with a ``TS.''  The growth rates of the
J and TS modes merge at the onset of oscillations
shown in the lower panel.
Dashed vertical lines demarcate the critical wavenumbers for growth of the
Jeans and two-stream modes.  The ``$+$'' and ``$-$'' signs in the frequency
plot denote the energy signature discussed in section 4.}
\label{f:quarticM2}
\end{figure}

The third-order resonance conditions for a triplet of modes are,
\beqa
m_{_1}k_{_1}+m_{_2}k_{_2}+m_{_3}k_{_3}&=&0 \nonumber\\
m_{_1}\omega_{_1}\pm m_{_2}\omega_{_2}\pm
m_{_3}\omega_{_3}&=&0\label{eq:rescon3}\\
\mid m_{_1}\mid+\mid m_{_2}\mid +\mid m_{_3}\mid
&=&3\hs\hs(m_{_1},m_{_2},m_{_3}\hspace{.15cm}{\rm integers}).\nonumber
\eeqa
which here may be satisfied by $(m_{_1},m_{_2},m_{_3})=(1,1,-1)$,
$(k_{_1},k_{_2},k_{_3})=(k_m,k_m,2k_m)$, and
$(\omega_{_1},\omega_{_2},\omega_{_3})=(\omega_+,\omega_-,\omega_-)$,
where the $\omega_{_j}$ are taken to be positive.  One can see
from figure \ref{f:quarticM2} that
$\omega_{_1}>\omega_{_2},\omega_{_3}$, so the relative signatures of
this triplet are immune to a Galilean shift.
Note from figure \ref{f:quarticM2}
that a resonant triplet involving two PEMs and
one NEM would not have robust relative signatures under a frame shift
since the PEMs have larger frequencies.

The lowest order nonlinear terms come from the third variation of $H$
expanded about the dynamical equilibrium,
\beq
\delta^3 H={\frac{1}{M}}\sum_{j=1}^{2}\int_{0}^{L}
dx\left({\frac{M^2}{2}}\delta\rho_{_j}\delta\phi_{_{jx}}^{2}+
{\frac{\gamma}{6}}\delta\rho_{_j}^{3}\right). \label{eq:hamnon}
\eeq
In terms of the Fourier amplitudes this expression is,
\beqa
\delta^3 H={\frac{L}{2}}\sum_{j=1}^{2}\sum_{{m,n=1}}^{\infty}
[-Mk_{_m}k_{_n}\left(\rho_{_{m+n}}^{(j)}\phi_{_{-m}}^{(j)}
\phi_{_{-n}}^{(j)}-\rho_{_{m-n}}^{(j)}\phi_{_{-m}}^{(j)}
\phi_{_n}^{(j)}+{\rm c.c.}\right)\label{eq:fourierham3}\\
+{\frac{\gamma}{3M}}\left(\rho_{_m}^{(j)}\rho_{_n}^{(j)}
\rho_{_{-m-n}}^{(j)}+\rho_{_m}^{(j)}\rho_{_{-n}}^{(j)}
\rho_{_{n-m}}^{(j)}+{\rm c.c.}\right)].\nonumber
\eeqa
We then effect the same transformations on (\ref{eq:fourierham3})
that led to the diagonalized free energy (\ref{eq:freeaa}).
This spawns a myriad of nonlinear terms, only some of which
survive an averaging process that leads to the Birkhoff normal form.$^{20}$

For a three-wave resonance, one finds after near-identity
transformations that the only higher
order terms contributing to the normal form are of the type,$^{14}$
\beq
{\cal O}(3){\rm\mskipp Terms}\sim J_{_1}^{|l|/2} J_{_2}^{|m|/2}
J_{_3}^{|n|/2} \sin\left( l\theta_{_1}+m\theta_{_2}+n\theta_{_3}\right),
\label{eq:cubicterm}
\eeq
with $|l|+|m|+|n|=3$.  The Hamiltonian up to $3^{\rm{rd}}$ order
terms for the resonant NEM/PEM triplets can be written as,
\beq
H=\omega_1 J_{_1}-\omega_2 J_{_2}-\omega_3 J_{_3}
+\alpha\sqrt{J_{_1}J_{_2}J_{_3}}\sin\left(\theta_{_1}+\theta_{_2}+\theta_{_3}\right),\label{eq:3waveham}
\eeq
where $\alpha=\alpha(k_m,M,L)$ is a nonlinear coupling constant that
is neither especially large or small in the parameter regime
of the three-wave resonance considered here.

Since the angles $(\theta_{_1},\theta_{_2},\theta_{_3})$
appear in only one combination in $H$, further simplification of
(\ref{eq:3waveham}) is possible via the generating function,
\beq
F_{_2}\left({\bf I},{\bf\theta}\right)={\frac{1}{2}}I_{_1}\left(
\theta_{_1}+\theta_{_2}\right)+I_{_2}\theta_{_2}+I_{_3}\theta_{_3},
\label{eq:genfunk}
\eeq
with ${\psi_{_j}}=${\large{${\frac{\partial F_{_2}}{\partial I_{_j}}}$}} and
${J_{_j}}=${\large{${\frac{\partial F_{_2}}{\partial \theta_{_j}}}$}}.
This canonical transformation yields,
\beq
H={\tilde\omega_{_1}}I_{_1}-\omega_{_3}I_{_3}+
{\frac{\alpha}{2}}\sqrt{I_{_1}\left(I_{_1}+2I_{_2}\right)I_{_3}}
\sin\left(2\psi_{_1}+\psi_{_3}\right),
\label{eq:3waveham2}
\eeq
with $2{\tilde\omega_{_1}}\equiv\omega_{_1}-\omega_{_2}$.
If one chooses
initial conditions satisfying $I_{_2}\equiv J_{_2}-J_{_1}=0$,
then $H$ is identical to the normal form of a two-wave interaction
originally presented by Cherry.$^{21,22}$
In terms of the $({\bf q},{\bf p})$ variables, Cherry's Hamiltonian is
\beq
H={\frac{1}{2}}{\tilde\omega_{_1}}\left(p_{_1}^2+q_{_1}^2\right)-
{\frac{1}{2}}{\omega_{_3}}\left(p_{_3}^2+q_{_3}^2\right)+
{\frac{\epsilon}{2}}
\left(2q_{_1} p_{_1} p_{_3}-q_{_3}\left[q_{_1}^2-p_{_1}^2\right]\right),
\label{eq:cherry}
\eeq
where $\epsilon=${\Large${\frac{\sqrt{2}\alpha}{4}}$}.

The dynamical system generated by the Cherry Hamiltonian is
integrable.  In the special case of a third-order resonance
with $\omega_{_3}=2{\tilde\omega_{_1}}$, there exists a
family of two-parameter solutions,
\beqa
q_{_1}={\frac{\sqrt{2}}{\xi-\epsilon
t}}\sin\left({\tilde\omega_{_1}}t+\eta\right)
\mskipp\mskipp&,&\mskipp\mskipp
p_{_1}=-{\frac{\sqrt{2}}{\xi-\epsilon
t}}\cos\left({\tilde\omega_{_1}}t+\eta\right)
\label{eq:rescherrysolns}\\
q_{_3}=-{\frac{1}{\xi-\epsilon t}}\sin\left(2{\tilde\omega_{_1}}t+2\eta\right)
\mskipp\mskipp&,&\mskipp\mskipp
p_{_3}=-{\frac{1}{\xi-\epsilon
    t}}\cos\left(2{\tilde\omega_{_1}}t+2\eta\right)\mskip,\nonumber
\eeqa
where $\xi$ and $\eta$ are constants depending on the initial
conditions.

The solutions (\ref{eq:rescherrysolns}) show the possibility of
{\it finite-time density singularities} when two negative energy
modes interact resonantly with a positive energy mode.
A system exhibiting this behavior is said to undergo
{\it explosive growth}, and it could
be an important mechanism for structure formation in galactic and
cosmological settings when relative motion between different fluid
species is involved.  If the resonance is detuned,
separatrices bounding stable orbits emerge in phase space, but
the dynamics are still prone to finite-amplitude
instability.

\section{DISSIPATIVE INSTABILITY}
\label{sec-dissipation}
\hs
We close our investigation of the consequences of negative energy
modes by examining the effects of dissipation on the linear
stability of the system.
With negative energy modes propagating through a dissipative medium,
we may expect new instabilities since the damping
can pump more negative energy into the wave.  This somewhat
counterintuitive effect of frictional forces in other contexts
was first pointed out by Kelvin and Tait$^{23}$
(see also Zajac$^{24}$).

Suppose that collisions are important at some stage in the development
of a gravitationally bound structure.
A simple model of this effect incorporates a dynamical friction
term $(-1)^j\nu\left({\bf u}_1-{\bf u}_2\right)$
on the right hand side of the momentum equations (\ref{eq:3Dmom}),
where $\nu$ is a positive damping coefficient.

In the dimensionless symmetric case the dispersion relation becomes,
\beqa
\omega^4-2i\nu\omega^3 +\left[1-2k^2(M^2+1)\right]\omega^2
\ns\ns&+&\ns\ns 2i\nu\left[k^2(M^2+1)-1\right]\omega\label{eq:dampeddis}\\
& &+k^2(M^2-1)\left[k^2(M^2-1)+1\right]=0. \nonumber
\eeqa
If we assume the damping is weak, $\nu\ll 1$, we may develop
(\ref{eq:dampeddis}) in a regular perturbation series
($\omega=\omega_{_0}+\nu\omega_{_1}+\ldots$)
to find the lowest order corrections to the frequencies
(\ref{eq:symroots}),
\beq
\omega_{_1}^{\pm}={\frac{i}{2}}\left(1\pm{\frac{1}{\sqrt{1-8k^2 M^2+16M^2
        k^4}}}\right), \label{eq:freqcorrection}
\eeq
where we assume $k^2\gg{\frac{M\pm\sqrt{M^2-1}}{4M}}$ to
avoid the singularity accompanying the vanishing denominator in
(\ref{eq:freqcorrection}) (see Casti$^{16}$ for the details).  
If we assume $M>1$ so that negative energy modes are present, then
a close examination of the corrections reveals that 
the dissipation promotes instability in the wavenumber band
${\frac{M\pm\sqrt{M^2-1}}{4M}}\ll k<{\frac{\sqrt{2}}{2}}$
for any $M\ge1$ {\it no matter how weak the damping}.  Since
the instability as $k^2\rightarrow{\frac{1}{2}}$ is realized only in the
$M\rightarrow\infty$ limit of the undamped problem, we see that
the dissipation indeed has the effect of destabilizing modes
that were stable in the conservative case.  A numerical
investigation revealed that this result holds for any $\nu>0$.

\begin{figure}[H]
\epsfxsize=4.5in
\centerline{\epsfbox{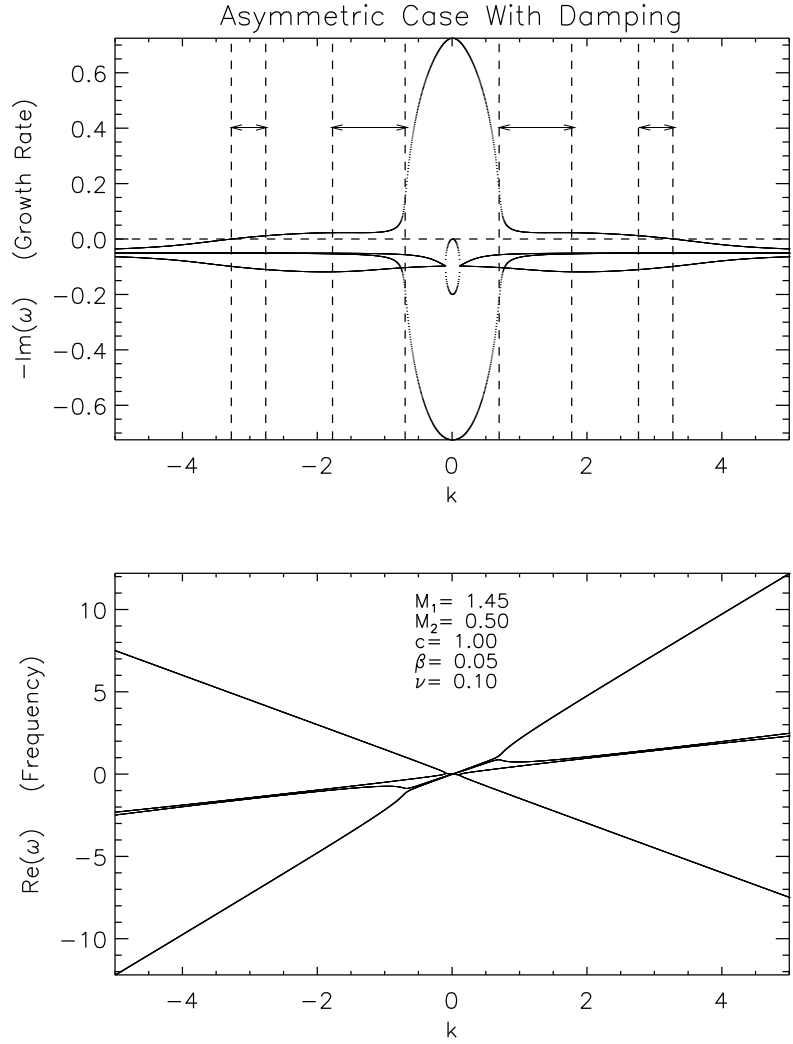}}
\caption{Dispersion curves for damped supersonic motion, $\nu=0.1$,
$M_{1}=1.45$, $M_{2}=0.5$, in the
asymmetric case with $\beta=.05$ and $c=1$.
The entire $k$-band separating the Jeans
bubble and two-stream bubbles of figure \ref{f:wispyM1.45} are now unstable.
The modal bands destabilized by the dissipation are demarcated by the
arrows between the dashed vertical lines.}
\label{f:damwispyM1.45}
\end{figure}

The modal bands destabilized by the damping become more significant
in the asymmetric case.  As remarked
in section \ref{sec-asymm},
bubbles of unstable two-stream modes can pinch off from the
Jeans-unstable bubble and result in well-separated instability
bands.  The inclusion
of dissipation can destabilize the entire band of modes separating
the bubbles, as well as some higher-$k$ modes beyond the undamped
two-stream bubble.  This is illustrated in figure
\ref{f:damwispyM1.45}.

One should not assume that any form of dissipation will destabilize
negative energy modes.  For instance, if each fluid feels only a drag
proportional to its own velocity, there are no new instabilities
even with relative motion.  In other words, the dissipation must
in some sense project onto the eigenspace spanned by the
negative energy modes in a way that decreases their energies.  This depends
not only on the nature of the dissipation, but also upon the 
initial equilibrium about which one perturbs.

The effect of damping can be understood 
by examining the time evolution of the free energy.
If the dissipation acts to increase the energy of a positive energy
mode or decrease the energy of a negative energy mode, then one
can show that pure imaginary eigenvalues take on a positive real
part.$^{25}$  To see that this is possible here, consider
the temporal change in the free energy, which in the symmetric case
can be written,
\beq
{\delta^2{\dot H}}=-\nu M\int_{0}^{L} dx\left(
\delta\phi_{{1x}}-\delta\phi_{{2x}}\right)^2
-\nu M\int_{0}^{L} dx\left(\delta\phi_1-\delta\phi_2\right)
\left(\delta\rho_{{1x}}+\delta\rho_{{2x}}\right).
\label{eq:hdot2}
\eeq
Since the first term of ${\delta^2{\dot H}}$ is negative definite,
the conditions for which
the free energy decays or grows in time is determined by the relative phasings
of the velocity and density perturbations comprising the second
term.  One may deduce the effect of the dissipation on any particular
mode of the conservative problem by inserting the undamped modes into
(\ref{eq:hdot2}), which yields a formula for ${\delta^2{\dot H}}$
valid up to ${\cal O}(\nu^2)$.
For subsonic relative motion, $M<1$, the expression (\ref{eq:hdot2}) 
is negative definite and the damping lives up to its name and causes the PEMs
to decay in time.  When $M>1$, ${\delta^2{\dot H}}$ can be either
positive or negative for an NEM depending on the value of $k$, which explains
why some NEMs are destabilized and others are damped in the usual
sense.

\section{DISCUSSION}
\label{sec-discussion}
\hs
The formation of structures through the action of gravity is much
analyzed in cosmology, galactic structure and cosmogony.
Most of this analysis is centered on the operation of gravitational
instability, though streaming fluids can resonantly interact via the
gravitational field to cause linear instability in spectral ranges
inaccessible to the traditional Jeans instability.  As
we have brought out here, the distinction between the two types of
unstable modes, the Jeans and the two-stream, becomes sharper when
one constituent is far denser than the other.  Much of the previous work
on the subject failed to take advantage of this crucial feature by
focusing attention on situations where each component exists in equal
abundance.  Even when the two-stream instability does not occur, if
the total energy of the gravitational two-stream interaction is
indefinite, the positive and negative energy modes that are
{\it stable} in the linear theory can interact to produce explosive
development of disturbances of arbitrarily small amplitude.  This
can be a significant aspect of the theory of structure formation.

There are many clear instances where the dynamics of interpenetrating
fluids may play a role in developing structures, but we close here
by suggesting that even when the streaming is not apparent, two-stream
dynamics may be relevant.  An interesting example is provided by the
coexistence of dark matter and luminous (baryonic) matter that is
generally believed to occur throughout the cosmos.  The locations of the
two kinds of matter seem to be well correlated, which would not be the
case if they were now streaming through each other.  On the other hand,
it might be reasonable to ask why there is this apparent correlation (or
anticorrelation in the case of negative gravitational density) of the two
kinds of material.   Even if they had once been in relative motion this
situation would not long persist, as we have seen.  But the outcome, as
far as large-scale structure is concerned, could be quite different if the
kinematic history of the interaction of the two matters had been richer
than has been supposed hitherto.  Given the indefiniteness of the
free energy if the initial streaming is large enough, waves of short
length scale could have interacted in an explosive manner to quickly
produce highly nonlinear density fluctuations.  This is a feature of
gravitational structure development that could be profitably studied,
particularly in situations where the background Hubble expansion
cannot be ignored.  The dynamics of a two-fluid system with
initially {\it time-dependent} relative motion is currently under
investigation.

\vspace{.5cm}
\begin{center}
{\bf REFERENCES}
\end{center}
\begin{flushleft}
\href 1.   Theys, J. C. \& E. A. Spiegel. 1977. Ap. J. {\bf 212:} 616.

\href 2.   Barnes, J. E. 1992.  Ap. J. {\bf 393:} 484.

\href 3.   Buneman, O. 1959.  Phys. Rev. {\bf 115:} 503.

\href 4.   Sweet, P. A. 1963.  Monthly Notices Roy. Astron. Soc. {\bf 125:}
285.

\href 5.  Lynden-Bell, D. 1967, in Relativity Theory and Astrophysics 2.
 Galactic Structure.

\hs\hs J. Ehlers, ed.  Providence: American Math. Soc.

\href 6.  Davidson, R. C. 1972.  Methods in Nonlinear Plasma Theory.
 Academic Press, 

\hs\hs New York.

\href 7.  Lovelace, R. V. E., K. P. Jore \& M. P. Haynes.  1997.  Ap. J.
{\bf 475:} 83.

\href 8.  Jeans, J. 1902.  Phil. Trans. R. Soc. {\bf 199A:} 49.

\href 9.  Ikeuchi, S., T. Nakamura \& F. Takahara.  1974.
Prog. Theor. Phys. {\bf 52:} 1807.

10.  Fridman, A. M. \& V. L. Polyachenko.  1984.  Physics of Gravitating
Systems II.

\hs\hs Springer-Verlag, New York.

11.  Araki, S. 1987.  Astron. J. {\bf 94:} 99.

12.  Spiegel, E. A. 1972, in Symposium on the Origin of the Solar
 System: 165.  H. Reeves,

\hs\hs ed.  Editions du Centre National de la
     Recherche Scientifique, Paris.

13.  de Carvalho, J. P. M. \& P. G. Macedo. 1995.  Astron. Astrophys.
{\bf 299:} 326.

14.  Kueny, C. S. \& P. J. Morrison. 1995.  Phys. Plasmas. {\bf 2:} 1926.

15.  Morrison, P. J. 1998.  Rev. Mod. Phys. {\bf 70:} 467.

16.  Casti, A. R. R. 1998.  Ph. D. Thesis.  Columbia University.

17.  Moser, J. K. 1958.  Comm. Pure Appl. Math. {\bf 11:} 81.

18.  Meyer, K. R. \& G. R. Hall. 1991.  Introduction to Hamiltonian
Dynamical Systems

\hs\hs and the N-Body Problem.  Springer-Verlag New York Inc.

19.  Weiland, J. \& H. Wilhelmsson. 1977.  Coherent Nonlinear
Interaction of Waves in

\hs\hs Plasmas.  Pergamon, Oxford.

20.  Ozorio de Almeida, A. M. 1988.  Hamiltonian Systems: Chaos and
Quantization.

\hs\hs Cambridge University Press, Cambridge.

21.  Cherry, T. M. 1925.  Trans. Cambridge Philos. Soc. {\bf 23:} 199.

22.  Whittaker, E. T. 1937.  Analytical Dynamics: 142.  Cambridge, London.

23.  Thompson, W. (Lord Kelvin) \& P. G. Tait. 1921.  Treatise on
Natural Philosophy, 

\hs\hs Part I:388.  Cambridge University Press, Cambridge.

24.  Zajac, E. E. 1964.  Journal of the Astronautical Sciences {\bf
  11:} 46.

25.  Mackay, R. S. 1991.  Phys. Lett. A. {\bf 155:} 266.

\end{flushleft}
\clearpage


\clearpage

\clearpage

\end{document}